\documentclass[aps,prl,twocolumn,superscriptaddress]{revtex4}
\usepackage{graphicx}
\usepackage{color}
\usepackage[tight]{units}
\usepackage{epstopdf}
\usepackage[normalem]{ulem}
\DeclareGraphicsExtensions{.pdf,.jpg,.png,.eps}
\usepackage{amsfonts}
\usepackage{amsmath}
\usepackage{amssymb}
\usepackage{color}
\usepackage{color}
\usepackage{lineno}
\usepackage{epstopdf}
\usepackage{adjustbox}

\usepackage[T1]{fontenc}
\usepackage{hyperref}

\begin{document}

\title{\textrm{Validating Mean Field Theory in a New Complex, Disordered High-Entropy Spinel Oxide}}

\author{Neha Sharma}
\affiliation{Department of Physics and Material Science, Thapar Institute of Engineering and Technology, Patiala 147004, India}

\author{Nikita Sharma}
\affiliation{Department of Physics and Material Science, Thapar Institute of Engineering and Technology, Patiala 147004, India}

\author{Jyoti Sharma}
\affiliation{UGC-DAE Consortium for Scientific Research, Khandwa Road, Indore 452001, Madhya Pradesh, India}
\author{S. D. Kaushik}
\affiliation{UGC-DAE Consortium for Scientific Research, BARC Campus, Trombay Mumbai - 400085, India}
\author{Sanjoy Kr. Mahatha}
\affiliation{UGC-DAE Consortium for Scientific Research, Khandwa Road, Indore 452001, Madhya Pradesh, India}
\author{Tirthankar Chakraborty}
\email[]{tirthankar@thapar.edu}
\affiliation{Department of Physics and Material Science, Thapar Institute of Engineering and Technology, Patiala 147004, India}
\author{Sourav Marik}
\email[]{soumarik@thapar.edu}
\affiliation{Department of Physics and Material Science, Thapar Institute of Engineering and Technology, Patiala 147004, India}

%\date{\today}

\begin{abstract}
The advent of novel high-entropy oxides has sparked substantial research interest due to their exceptional functional properties, which often surpass the mere sum of their constituent elements' characteristics. This study introduces a complex high-entropy spinel oxide with composition (Ni$_{0.2}$Mg$_{0.2}$Co$_{0.2}$Cu$_{0.2}$Zn$_{0.2}$)(Mn$_{0.66}$Fe$_{0.66}$Cr$_{0.66}$)O$_{4}$. We performed comprehensive structural (X-ray and Neutron diffraction), microstructural, magnetic, and local electronic structure investigations on this material. Despite the material's high degree of disorder, detailed magnetization measurements and low temperature neutron powder diffraction studies reveal long-range ferrimagnetic ordering beginning at 293 K. The sample exhibits a high saturation magnetization of 766 emu-cm${^3}$ (at 50 K), a low coercivity (H$_C$) of 100 Oe (50 K), a high transition temperature (T$_C$) around room temperature, and high resistivity value of 4000 Ohm-cm at room temperature, 
indicating its potential for high density memory devices. The magnetic structure is determined using a collinear-type ferrimagnetic model with a propagation vector k = 0,0,0. 
Various analytical techniques, including modified Arrott plots, Kouvel-Fischer analysis, and critical isotherm analysis, are employed to investigate the phase transitions and magnetic properties of this complex system. Our results indicate a second-order phase transition. Remarkably, despite the complex structure and significant disorder, the critical exponents obtained are consistent with the mean field model. The high entropy leads to a remarkably homogeneous distribution of multiple cations, validating the approximation of average local magnetic environments and supporting the mean field theory.

\end{abstract}

\maketitle

\section{Introduction}
\label{S:1}
    High entropy oxides (HEOs) have recently emerged as a promising and exciting field of research due to their extraordinary properties and potential for revolutionizing of material design and properties \cite{oses2020high, aamlid2023understanding, miracle2017critical, ye2016high}. These compounds are characterized by incorporating multiple elements, typically five or more in equimolar ratios, into a single crystallographic site. This incorporation results in enhanced configurational entropy ($\Delta S_{conf}$) which stabilizes the structure of these compounds, creating what is known as an entropy-stabilized phase. The first high entropy oxide, (Mg$_{0.2}$Co$_{0.2}$Ni$_{0.2}$Cu$_{0.2}$Zn$_{0.2}$)O was stabilized and investigated by Rost et al. \cite{rost2015entropy} in 2015. This pioneering work paved the way for extensive research into high entropy oxides (HEOs), driven by their easy synthesis, remarkable stability, exceptional functional properties, and the unique design flexibility that includes multiple cations in these materials.

    HEOs exhibit a range of exceptional properties that make them promising for various technological applications, including catalysts, thermoelectrics, and magnetic materials \cite{yan2020functional, sarkar2019high, anandkumar2023synthesis, jana2023designing, chen2024b, katzbaer2023band, cocconcelli2024spin, jin2023robust, mazza2022designing}. Among HEOs, the spinel family stands out as an intriguing material system for studying the interaction between disorder-driven complexity and the relationship between structural and functional properties. In the spinel structure AB$_2$O$_4$, A ions are located at tetrahedral sites, while B ions occupy octahedral sites. The B ions arrange themselves into a pyrochlore network of interconnected tetrahedra, which leads to substantial magnetic frustration. Concurrently, the A cations form a diamond-like network that exhibits antiferromagnetic collinear order, adding to the magnetic frustration \cite{buttgen2004orbital,fritsch2004spin}. The unique crystal structure of the spinel system allows for entropy modification at both A and B sites, providing a rich platform for exploring tunable properties in spintronics and magnetism. Prior research on high entropy spinel oxides has revealed intriguing physical and magnetic characteristics, including tunable magnetism, robust ferrimagnetism, switchable magnetic anisotropy, enhanced exchange bias effect, and high magnetic frustration \cite{johnstone2022entropy, Robustmag, sharma2023large, sharma2024complex, sarkar2023high}. The high entropy stabilization facilitates precise atomic-scale tuning of materials, enabling the development of materials with tailored properties for specific applications. However, synthesizing HEOs with precise compositions and desired properties remains a significant challenge. This creates the need for detailed exploration and investigation of new strongly correlated high entropy oxides with novel structures and intriguing properties. To date, most studies have focused on the structure and physicochemical properties of new high entropy oxides, while comprehensive studies on the homogeneity of the magnetic structure (despite high disorder) and the nature of magnetic phase transitions through critical analysis are still missing.
    
    Critical analysis plays a crucial role in understanding the phase transitions and magnetic properties of complex systems. The investigation of critical behavior at the ferromagnetic to paramagnetic phase transition involves studying critical exponents using various analytical techniques such as modified Arrott plots, Kouvel-Fischer, and critical isotherm analysis. These analyses reveal the nature of the magnetic phase transition elucidating the conventional universality classes. In this paper, we studied the critical behavior of a new high entropy spinel oxide system with the chemical composition (Ni$_{0.2}$Mg$_{0.2}$Co$_{0.2}$Cu$_{0.2}$Zn$_{0.2}$)(Mn$_{0.666}$Fe$_{0.666}$Cr$_{0.666}$)O$_{4}$. Notably, magnetic cations are incorporated in both A and B sites. The nuclear and magnetic structure of the material is characterized using X-ray and neutron powder diffraction measurements. Through various approaches, we performed an in-depth analysis of the critical exponents around phase transition temperature to investigate and elucidate the nature of the magnetic phase transition. Remarkably, despite the high degree of disorder, the derived critical exponents closely match those predicted by the mean field model of conventional universality classes. We propose that the high configurational entropy in this complex oxide may contribute to its conformity to such a simplified model.

\begin{figure*}
\includegraphics[width=0.75\textheight]{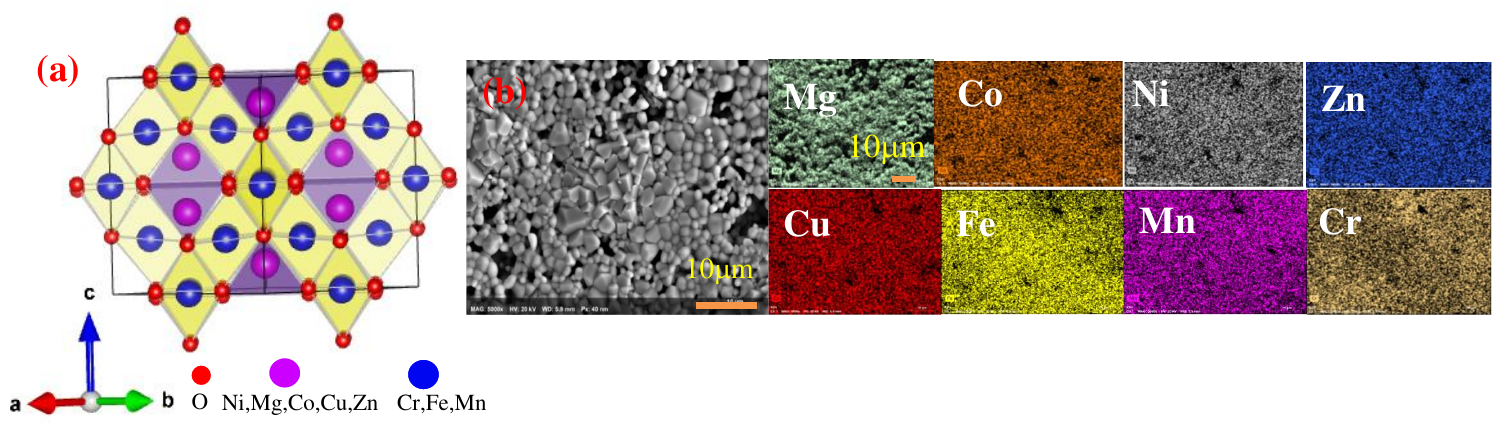}
\caption{\label{Figure 1:STR} (a) Crystal structure of the the material (AB$_2$O$_4$ type spinel). (b) Room temperature FESEM image and corresponding EDS mapping of (Ni$_{0.2}$Mg$_{0.2}$Co$_{0.2}$Cu$_{0.2}$Zn$_{0.2}$)(Mn$_{0.66}$Fe$_{0.66}$Cr$_{0.66}$)O$_{4}$.}
\end{figure*}

\section{Experimental Details}
\textbf{Synthesis.}
The polycrystalline material having chemical composition (Ni$_{0.2}$Mg$_{0.2}$Co$_{0.2}$Cu$_{0.2}$Zn$_{0.2}$)(Mn$_{0.666}$Fe$_{0.666}$Cr$_{0.666}$)O$_{4}$ are prepared via the conventional solid-state reaction method.The compound was produced using stoichiometric amounts of high purity powders, specifically Co$_3$O$_4$ (99.9\%), CuO (99.9\%), ZnO (99.9\%), NiO (99.9\%) and Cr$_2$O$_3$ (99.9\%),, MgO(99.99\%) and MnO$_2$ (99.9\%) and Fe$_2$O$_3$ (99.99\%). Using a mortar and pestle, the components were first combined. After the mixture was homogenized, it underwent multiple heat treatments before being air quenched and then final sintering of material was performed at 1373 K for 36 hours.

\textbf{X-ray Diffraction and Neutron Diffraction.}
We have performed X-ray diffraction (XRD) experiments using the powder of the materials. The powder XRD patterns were collected at room temperature (RT) using a Rigaku X-ray diffractometer (Cu-K$_\alpha$, $\lambda$ = 1.54056 \AA). Further, Neutron diffraction (NPD) measurements were conducted on (Ni$_{0.2}$Mg$_{0.2}$Co$_{0.2}$Cu$_{0.2}$Zn$_{0.2}$)(Mn$_{0.666}$Fe$_{0.666}$Cr$_{0.666}$)O$_{4}$  at 10 K and 300 K using a multi-Position Sensitive Detector (PSD) Focusing Crystal Diffractometer (FCD) provided by the UGC-DAE Consortium for Scientific Research at the Mumbai center, located at the National Facility for Neutron Beam Research (NFNBR) within the Dhruva reactor in Mumbai, India. These measurements were performed at a wavelength of 1.48 Å. We used a vanadium sample holder, which was directly exposed to the neutron beam. The diffracted neutrons were recorded using He${^3}$ filled linear position-sensitive detectors, employing the charge division method. 

\textbf{Field Emission Scanning Electron Microscopy (FESEM), and Energy Dispersive Spectroscopy (EDS).} 
We utilized field emission scanning electron microscopy (FE-SEM) and electron diffraction to investigate the microstructure and chemical homogeneity of the high entropy materials. The chemical composition was determined using Energy Dispersive X-ray Spectroscopy (EDS) with a BRUKER XFlash 6160 detector, while the microstructure analysis was conducted with a ZEISS GEMINI FE-SEM.

\textbf{Magnetic Measurements.}
Magnetization measurements were conducted using a Quantum Design MPMS 3 superconducting quantum interference device (SQUID). Both field-cooled (FC) and zero-field-cooled (ZFC) modes were employed across a temperature range from 4 K to 375 K. In addition, multiple M-H loops were measured at temperatures ranging from 4 K to 310 K, utilizing magnetic fields varying between 100 Oe and 2 T.

\textbf{X-Ray Absorption Spectroscopy.} Soft X-ray absorption spectroscopy (SXAS) measurements on the L-edges of the constituent transition metal atoms were performed at the SXAS beamline (BL-01) of the Indus2 synchrotron source, located at the Raja Ramanna Centre for Advanced Technology (RRCAT) in Indore, India. All the measurements were conducted at room temperature with an energy resolution of approximately 0.5 eV. The CTM4XAS (Charge transfer multiplet program for X-ray absorption spectroscopy) software was used to simulate the XAS spectra of various transition metal elements in the sample. 

\section{Scaling analysis}
The scaling hypothesis and analysis are essential in understanding critical phenomena, where physical properties remain scale-invariant near the critical point. The hypothesis states that certain physical quantities, which undergo changes during the phase transition, can be described as functions of the external variables driving the transition. Scaling analysis involves extracting critical exponents from data collapses, facilitating the classification of transitions based on universal critical behavior that transcends the microscopic details of individual systems. According to the scaling hypothesis, the spontaneous magnetization 
$M_S(T)$ below the phase transition temperature ($T_c$), the inverse magnetic susceptibility $\chi_0^{-1}(T)$ above $T_c$ and the magnetization at $T_c$
should exhibit the following relationships for a second-order magnetic phase transition \cite{stanley1971phase}:
	\begin{equation}
	M_S(T)=M_0 (-\epsilon)^\beta, \hspace{1cm} \epsilon<0
		\label{eqn_1}
	\end{equation}
	\begin{equation}
	\chi_0^{-1}(T)=\Gamma (\epsilon)^\gamma, \hspace{1cm} \epsilon>0
	\label{eqn_2}
	\end{equation}
	\begin{equation}
	M=X H^{\frac{1}{\delta}}, \hspace{1cm} \epsilon=0
	\label{eqn_3}
	\end{equation}
where, $M_0, \Gamma$ and $X$ are the critical amplitudes and $\beta, \gamma$ and $\delta$ are the critical exponents. $\epsilon$ is the reduced temperature defined as $\epsilon=\left( \frac{T-T_c}{T_c} \right)$.

Scaling theory provides a method to relate the critical exponents $\beta$ and $\gamma$ \cite{cardy1996scaling, pathria2011edition}, resulting in the formulation of the magnetic equation of state, which can be expressed as follows \cite{kaul1985static}:
	\begin{equation}
  	M(H,\epsilon)=\epsilon^\beta f_\pm \left(\frac{H}{\epsilon^{\beta+\gamma}}\right)
  	\label{eqn_4}
  	\end{equation}
	 where, $f_+(T>T_c)$ and $f_-(T<T_c)$ are the regular analytical functions. 
    By introducing renormalized magnetization 
	$m\equiv\epsilon^{-\beta}M(H,\epsilon)$ and renormalized field $h\equiv\epsilon^{-(\beta+\gamma)}H$, Equation \ref{eqn_4} can be simplified as follows:
	\begin{equation}
	m=f_{\pm}(h)
	\label{eqn_5}
	\end{equation}
 According to Equation \ref{eqn_5}, scaling relations implies that when plotting $m$ against $h$, the data should form distinctive patterns known as universal curves which consist of two separate branches: one for temperatures above $T_c$ where $\epsilon$ is positive, and the other for temperatures below $T_c$ where $\epsilon$ is negative \cite{arrott1957criterion}. The emergence of these universal curves confirms the validity of the scaling relations. Furthermore, they are essential for accurately determining the critical exponents $\beta$, $\gamma$, and $\delta$ within the critical regime.

\begin{figure}
\centering
\includegraphics[width=1.0\columnwidth]{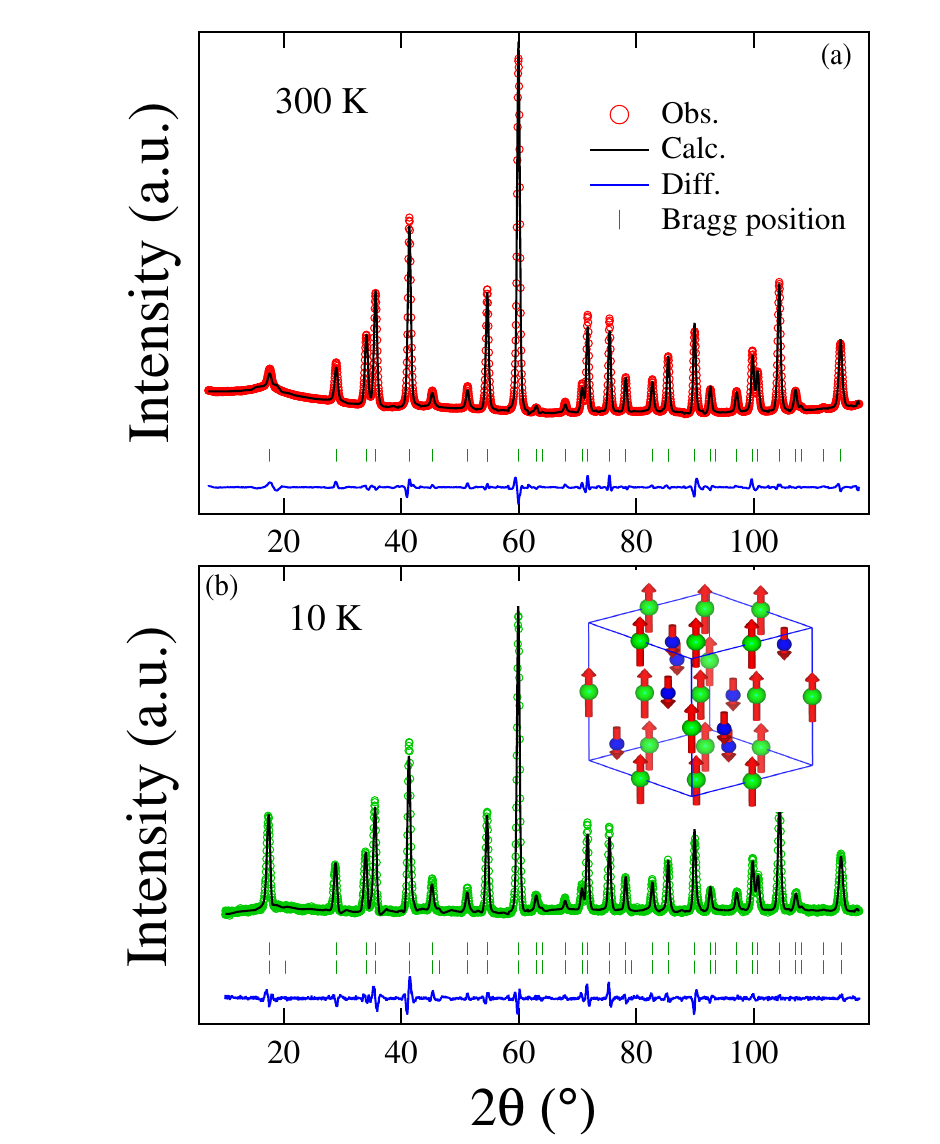}
\caption{\label{Fig2}
Rietveld refinement patterns of the  (a) 300 K and (b) 10 K neutron powder diffraction (NPD) patterns for (Ni$_{0.2}$Mg$_{0.2}$Co$_{0.2}$Cu$_{0.2}$Zn$_{0.2}$)(Mn$_{0.66}$Fe$_{0.66}$Cr$_{0.66}$)O$_{4}$. Lower ticks in figure (b) shows the magnetic peak positions with k = 0, 0, 0. The obtained magnetic structure is shown as an inset in figure (b), green spheres highlight the B - site cations and blue spheres highlight the A - site cations.}
	\end{figure}

\section{Results and Discussions}

\subsection{Structure and Magnetic Structure}
 Phase purity and compositional analysis of the sample are verified with the Rietveld refinement of the room temperature powder X-ray diffraction pattern (RT-XRD) and Energy-dispersive X-ray spectroscopy (EDS) mapping. The RT-XRD pattern of (Ni$_{0.2}$Mg$_{0.2}$Co$_{0.2}$Cu$_{0.2}$Zn$_{0.2}$)(Mn$_{0.66}$Fe$_{0.66}$Cr$_{0.66}$)O$_{4}$  reveals that the sample crystallizes in a cubic structure with the space group $Fd\overline{3}m$. Figure 1 (b) presents the ambient temperature field emission scanning electron microscopy (FESEM) image and EDS elemental mapping results. Ambient temperature EDS mapping confirms the stoichiometric chemical composition, indicating a microscale homogeneous distribution of all elements across the sample. To further investigate the detailed nuclear and magnetic structures, neutron powder diffraction (NPD) patterns were collected at 10 K and 300 K. The results from NPD Rietveld refinements are utilized to elucidate the crystal structure and magnetic structure of (Ni$_{0.2}$Mg$_{0.2}$Co$_{0.2}$Cu$_{0.2}$Zn$_{0.2}$)(Mn$_{0.66}$Fe$_{0.66}$Cr$_{0.66}$)O$_{4}$. Figure 2 presents the neutron powder diffraction (NPD) Rietveld refinement plots (10 K and 300 K) of the sample. The structural parameters derived from the 10 K and 300 K NPD refinements for (Ni$_{0.2}$Mg$_{0.2}$Co$_{0.2}$Cu$_{0.2}$Zn$_{0.2}$)(Mn$_{0.66}$Fe$_{0.66}$Cr$_{0.66}$)O$_{4}$ are summarized in Table 1. The crystal structure is depicted in Figure 1 (a). Here, metal cations occupy both octahedral and tetrahedral coordination environments, surrounded by oxygen atoms, forming two distinct sets of magnetic sublattices. Typically, B cations (Cr, Mn, and Fe for the present sample) occupy octahedral sites, configuring a lattice similar to pyrochlore, which induces highly frustrated magnetic interactions. Conversely, A cations (Ni, Mg, Co, Cu, and Zn for the present sample) occupy tetrahedral sites, creating an eightfold coordination reminiscent of a diamond lattice structure.

Refinement of oxygen occupancy shows no deviation from full occupancy. The cations Co, Mg, Cu, and Zn exclusively occupy the tetrahedral A sites (8a (0.125, 0.125, 0.125)), while Fe, Cr, and Mn are located in the octahedral B sites (16d (0.5, 0.5, 0.5)). Ni and Mn cations are found in both crystallographic sites, with 40\% of the Ni cations residing in the octahedral sites. Despite the high disorder and chemical complexity, the low temperature (10 K) neutron powder diffraction (NPD) data reveal long-range magnetic ordering. All peaks with magnetic origins align with the allowed nuclear reflections, and the magnetic structure is described using the propagation vector k = (0,0,0). The NPD magnetic reflections for long-range magnetic order fit best with a collinear ferrimagnetic ordering model, similar to those proposed for CoFe$_2$O$_4$ and CoFeRhO$_4$ \cite{CoFe2O4MagneticStr, PhysRevMaterialsCoFeRh}. The magnetic structure is shown in the inset of figure 2(b). The absence of the (200) magnetic peak suggests that a non-collinear spin arrangement is unlikely. According to the Yafet-Kittel model, the presence of the (200) magnetic Bragg peak would indicate long-range spin canting, involving the transverse component ordering of the magnetic spins. However, this peak is not observed at 20.35 $\deg$, while the (400) peak is clearly seen at 2$\theta$ = 41.39 $\deg$. The magnetic moments in the tetrahedral A sites (M(T$_d$)) and octahedral B sites (M(Oct)) are detailed in Table 1. We have observed a discrepancy between the experimentally measured and theoretically calculated magnetic moments for these sites. The observed reduction in magnetic moments can be attributed to the high disorder and local canting of spins. The extremely complex cationic distribution and the presence of nonmagnetic cations (Mg and Zn) within the structure could result in competing magnetic interactions, leading to nonuniform local spin canting. This reduction in magnetic moments due to local spin canting has also been observed in CoFe$_2$O$_4$ and Ti-doped CoFe$_2$O$_4$ \cite{CoFe2O4MagneticStr, TidopedCoFe}. Nevertheless, both materials are suggested to exhibit collinear ferrimagnetic long-range magnetic ordering.

 \begin{table*}
\small
  \caption{\ Atomic and structural parameters as obtained from the 10 K and 300 K NPD refinements for (Ni$_{0.2}$Mg$_{0.2}$Co$_{0.2}$Cu$_{0.2}$Zn$_{0.2}$)(Mn$_{0.66}$Fe$_{0.66}$Cr$_{0.66}$)O$_{4}$.}
  \label{tbl 2:NPD}
  \begin{tabular*}{\textwidth}{@{\extracolsep{\fill}}lll}
    \hline
      & 300 K & 10 K\\
    \hline
    Space Group & $Fd\overline{3}m$ & $Fd\overline{3}m$\\

    a = b = c (\AA) & 8.3804 (1) \AA  & 8.3760 (1) \AA\\
    Ni (0.125, 0.125, 0.125) &  & \\
    Occupancy & 0.08 (2) & 0.08 (2)\\
    B$_{iso}$ & 0.60 (2) & 0.41 (2)\\
     Mn (0.125, 0.125, 0.125) &  & \\
    Occupancy & 0.12 (2) & 0.12 (2)\\
    B$_{iso}$ & 0.60 (2) & 0.41 (2)\\
    Mg (0.125, 0.125, 0.125) &  & \\
    Occupancy & 0.2  & 0.2\\
    B$_{iso}$ & 0.60 (2) & 0.41 (2)\\
     Co (0.125, 0.125, 0.125) &  & \\
    Occupancy & 0.2 & 0.2\\
    B$_{iso}$ & 0.60 (2) & 0.41 (2)\\
     Cu (0.125, 0.125, 0.125) &  & \\
    Occupancy & 0.2 & 0.2\\
    B$_{iso}$ & 0.60 (2) & 0.41 (2)\\
     Zn (0.125, 0.125, 0.125) &  & \\
    Occupancy & 0.2 & 0.2\\
    B$_{iso}$ & 0.60 (2) & 0.41 (2)\\
    Fe (0.5, 0.5, 0.5) &  & \\
    Occupancy & 0.66 & 0.66\\
    B$_{iso}$ & 0.39 (2) & 0.10 (1)\\
    Cr (0.5, 0.5, 0.5) &  & \\
    Occupancy & 0.66 & 0.66\\
    B$_{iso}$ & 0.39 & 0.10 (1)\\
     Mn (0.5, 0.5, 0.5) &  & \\
    Occupancy & 0.52 (2) & 0.52 (2)\\
    B$_{iso}$ & 0.39 (2) & 0.10 (1)\\
     O (x,x,x) &  & \\
    x = 0.2608 (1) & 0.2608 (1)\\
    Occupancy & 1.0 & 1.0\\
    B$_{iso}$ & 0.50 (1) & 0.35 (1)\\
    B - B distance & 2.9629 (1) \AA & 2.9614 (1) \AA\\
    A - B & 3.4743 (1) \AA & 3.4725 (1) \AA\\
     $\chi^2$ & 9.5 & 3.73\\
      R$_{P}$ & 8.25 & 3.65 \\
        R$_{WP}$ & 7.60 & 4.92 \\
        R$_{Mag}$ & - & 2.92 \\
         M (T$_d$) & - & 2.2 (1) $\mu$$_B$ \\
          M (Oct.) & - & 3.3 (1) $\mu$$_B$ \\ \\
    \hline
  \end{tabular*}
\end{table*}

 \begin{figure}
\includegraphics[width=0.35\textheight]{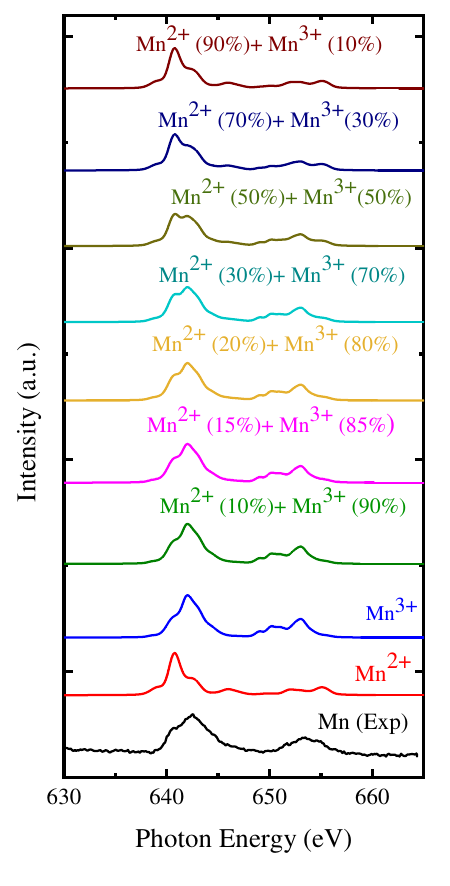}
\caption{\label{XASMn}
RT - XAS data and simulated Mn L-edge of the 
(Ni$_{0.2}$Mg$_{0.2}$Co$_{0.2}$Cu$_{0.2}$Zn$_{0.2}$)(Mn$_{0.66}$Fe$_{0.66}$Cr$_{0.66}$)O$_{4}$.}
	\end{figure}

 \begin{figure*}
\includegraphics[width=0.70\textheight]{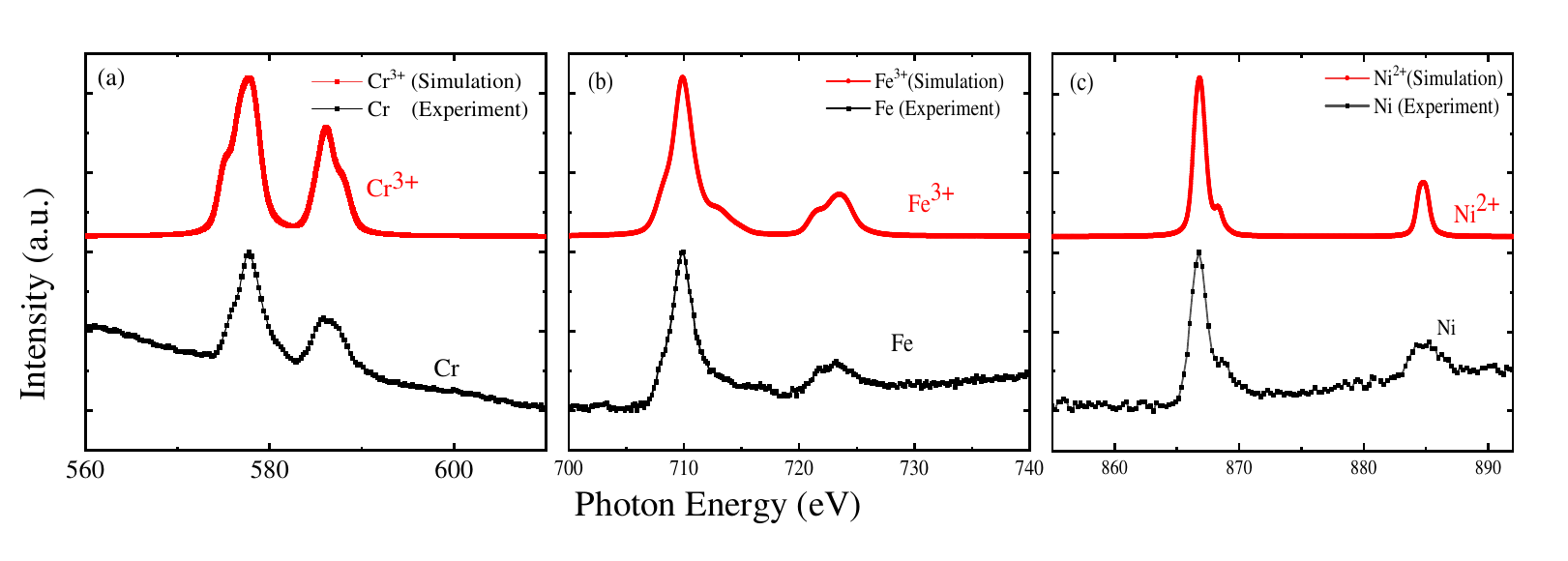}
\caption{\label{XASall}
The simulated (red) and experimental (black) XAS spectra of (a) Ni L$_{2,3}$, (b) Fe L$_{2,3}$ and
(c) Cr L$_{2,3}$ edges of
(Ni$_{0.2}$Mg$_{0.2}$Co$_{0.2}$Cu$_{0.2}$Zn$_{0.2}$)(Mn$_{0.66}$Fe$_{0.66}$Cr$_{0.66}$)O$_{4}$..}
	\end{figure*}

 \begin{figure}
\includegraphics[width=0.35\textheight]{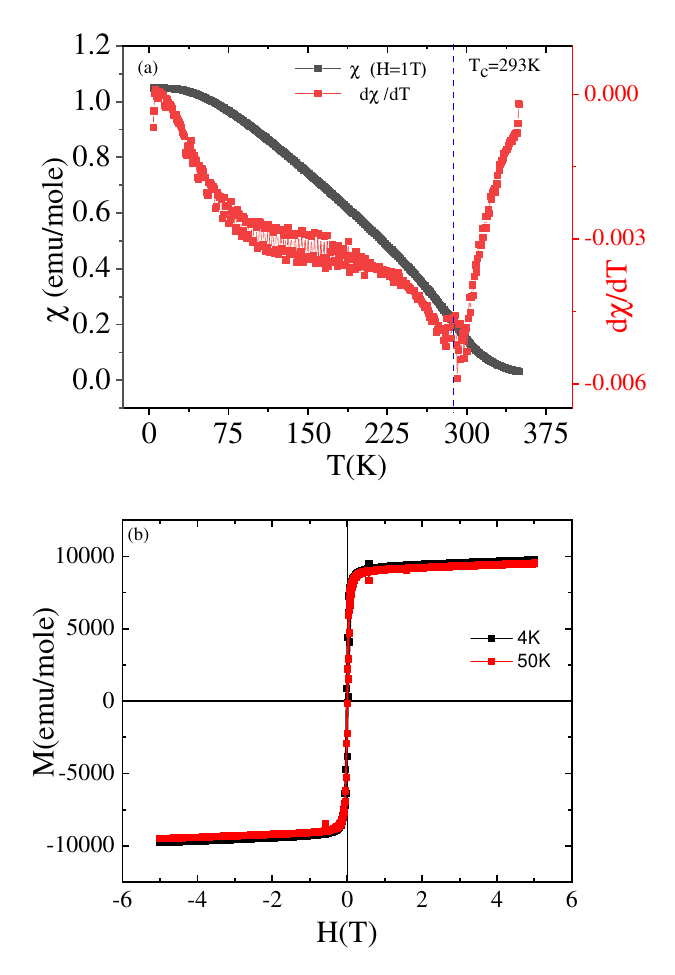}
\caption{\label{Fig3-Chi Vs T}
(a) Temperature dependence of magnetic susceptibility measured under a field-cooled (FC) protocol with an external field of 1 T. The derivative of the susceptibility is also shown, indicating the transition occurring around 293 K (marked by the dashed vertical line). (b) Magnetic field dependence of magnetization measured at 4 K and 50 K.}
	\end{figure}
 
 \subsection{X-ray absorption spectroscopy}
X-ray Absorption Spectroscopy (XAS) is the measurement of transition of core level electron to the states near or above the Fermi level in order to investigate the electronic structure and local atomic environment of the absorbing atom. The energy position and shape of the XAS spectra provide vivid information about the oxidation state, coordination number, and interatomic distances in a manner that is element-sensitive. Figure 3 displays the experimental XAS spectra of the Mn L-edge along with the simulated spectra of Mn$^{3+}$ in octahedral geometry and Mn$^{2+}$ in tetrahedral geometry. In order to get an estimate of relative contributions of octahedral-Mn$^{3+}$ and tetrahedral-Mn$^{2+}$ in the studied sample, the simulated spectra of Mn$^{3+}$ and Mn$^{2+}$ are mixed in different proportions and the resultant spectra are compared with the experimental spectra. We found the resultant simulated spectra with 15\% Mn$^{2+}$ and 85\% Mn$^{3+}$ matches well with the experimental one. The simulated spectra were
generated using the CTM4XAS software package. The software comprises three theoretical components that corresponds to the atomic multiplet theory, crystal field theory, and charge transfer theory respectively. It allows to model and interpret the electronic structure and local atomic environment of materials by providing detailed simulations of spectral features. For Mn$^{3+}$ (and 2+), the parameters were set as follows: 10Dq = 1.2 (1.3), Fdd = 1.5 (0.8), Fpd = 1.5 (0.8), Gpd = 1.5 (0.8), Lorentzian broadening = 0.2, and Gaussian broadening = 0.4. Similarly, the L-edge XAS spectra of Ni, Fe, and Cr have been simulated and shown in comparison with the experimental spectra in Figure 4 (a-c). The Ni XAS spectra (Figure 4(a)) indicate the presence of Ni$^{2+}$ oxidation state. However, the asymmetry in the peak shape suggests that some Ni may occupy octahedral positions, possibly exchanging places with Mn$^{2+}$ in the tetrahedral sites. This trend is also observed in our NPD refinements (table 1). The Cr XAS spectra confirm a Cr$^{3+}$ oxidation state. Similarly, the Fe L$_{2,3}$-edge spectra (Figure 4(c)) are consistent with Fe$^{3+}$ oxidation state. Based on our NPD and RT-XAS analysis, we found that the chemical formula of the sample can be written as (Ni$_{0.1}$Mn$_{0.1}$Mg$_{0.2}$Co$_{0.2}$Cu$_{0.2}$Zn$_{0.2}$)(Mn$_{0.56}$Ni$_{0.1}$Fe$_{0.66}$Cr$_{0.66}$)O$_{4}$. Nevertheless, we will refer to our sample as (Ni$_{0.2}$Mg$_{0.2}$Co$_{0.2}$Cu$_{0.2}$Zn$_{0.2}$)(Mn$_{0.66}$Fe$_{0.66}$Cr$_{0.66}$)O$_{4}$ in the rest of the manuscript.

 \subsection{Magnetism and Arrott plot}
  Figure 5 (a) shows the temperature dependence of magnetic susceptibility ($\chi$) measured under a field-cooled (FC) protocol with an external field of 1 T, within temperature range of 2 K to 370 K. As the temperature decreases, $\chi$ begins to increase sharply below 300 K, indicating the onset of long-range magnetic ferrimagnetic (FIM) ordering. The transition onset temperature is determined from the derivative of the curve ($d\chi/dT$), which exhibits a dip around 293 K, identifying this as the Curie temperature ($T_c$). The magnetization measurements as a function of temperature (Figure 5 (b)) reveal a sharp magnetization saturation with a very low coercivity of 100 Oe at 50 K. It highlights a remarkably high saturation magnetization (M$_S$) of 766 emu cm$^3$ at 50 K. Resistivity measurements demonstrate an impressive resistivity of 4000 ohm-cm at room temperature (Figure 1 in the supporting information file \cite{SI}). Additionally, the sample reaches magnetic saturation under a very weak field of 1500 Oe, highlighting its potential as a soft magnetic insulator material.
  
  To investigate the nature of the phase transition in detail, we conducted a critical analysis as described below. In conventional practices, critical exponents and critical temperature are initially derived from Arrott plots assuming mean field interactions within the system which is characterized by $\beta=0.5$ and $\gamma=1$. Thus, an Arrott plot consists of plotting isotherms as $M^2$ versus $H/M$ which, in ideal situation, yields a series of parallel straight lines near $T_c$, with the line at $T=T_c$ passing through the origin.
 From this plot, $\chi_0^{-1}(T)$ and $M_S(T)$ can be determined from the intercepts of the curves on positive $H/M$ and $M^2$ axes respectively.

\begin{figure*}[!t]
	\centering
	\includegraphics[width=0.85\textheight]{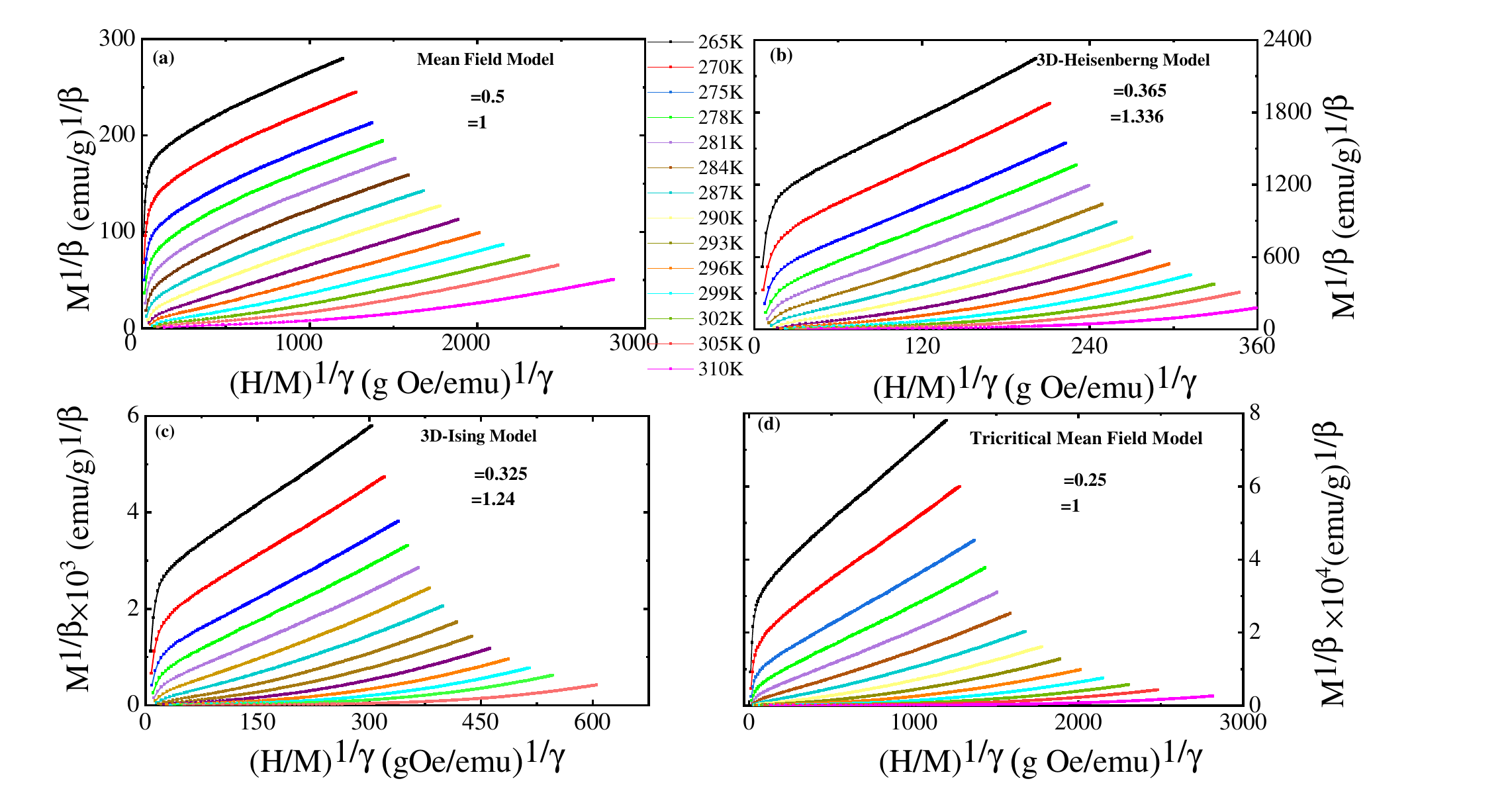}
	\caption{\label{Fig 4 All models } $M^{1/\beta}$ vs. $(H/M)^{1/\gamma}$ showing (a) Arrott plots considering mean field model ($\beta=0.5, \gamma=1$) and modified Arrott plots considering (b) 3D-Ising model ($\beta=0.325, \gamma=1.24$), (c) tricritical mean field model ($\beta=0.25, \gamma=1$), (d) 3D-Heisenberg model ($\beta=0.365, \gamma=1.336$). }
	\end{figure*}

 Figure 6 displays the Arrott plot around $T_c$ for this system. Except the one with mean field model (Figure \ref{Fig 4 All models } (a)), the nonlinear nature of the curves, even at high fields, indicates that these values of the critical exponents are not appropriate. Additionally, the order of the magnetic phase transition can be determined from the slope of the curves using the Banerjee criterion \cite{banerjee1964generalised}. The positive slope of the curves confirms that the phase transition is of second order.
 
 Even though the mean field model produces reasonably parallel straight lines in the higher field region, more accurate values of the exponents are rigorously investigated beyond these approximations to gain deeper insights into the microscopic nature of second-order phase transitions. A modified Arrott plot, known as the Arrott-Noakes equation, is employed as follows:
    \begin{equation}
	\left(\frac{H}{M}\right)^\gamma = a \frac{(T-T_c)}{T}+bM^{\frac{1}{\beta}}
	\label{eqn_6}
	\end{equation}
 where, $a$ and $b$ are constants. 
 This approach represents the most general form of the equation of state near the critical point, adhering to the scaling laws. Within this magnetic equation of state, different theoretical models correspond to distinct sets of critical exponents, such as the 3D-Ising model ($\beta=0.325$, $\gamma=1.24$), the tricritical mean field model ($\beta=0.25$, $\gamma=1$), and the 3D-Heisenberg model ($\beta=0.365$, $\gamma=1.336$) \cite{pathria2011edition, le1977critical, pelissetto2002critical}.
 Modified Arrott plots according to these models are shown in Figure \ref{Fig 4 All models }(b), (c), and (d) respectively which reveals curved lines instead of straight lines. Notably, none of the plots exhibit a line passing through the origin near the $T_C$ of 293 K. Consequently, these observations implies that these conventional models are not appropriate to explain the behavior of our systems definitively. This necessities further modification of the critical exponents in order to get an optimal equation of state that explains the critical behavior.

 Determining a suitable set of exponents that would yield parallel straight lines, as described by Equation \ref{eqn_6}, is challenging since it depends on two free parameters, $\beta$ and $\gamma$. To avoid fitting inaccuracies and systematic errors, we adopted a rigorous iterative approach to determine the values of $\beta$ and $\gamma$. This involves utilizing a modified Arrott plot (Equation \ref{eqn_6}) with initial estimates of $\beta$ and $\gamma$, derived from intercepts of linear extrapolations of isotherms in the high-field region on the $M^{1/\beta}$ and $(H/M)^{1/\gamma}$ axes, respectively. These estimates of $M_S(T)$ and $\chi_0^{-1}(T)$ are then employed to fit Equations \ref{eqn_1} and \ref{eqn_2}. From Equation \ref{eqn_1}, the slope of the straight-line fit of $\log[M_S(T)]$ versus $\log(\epsilon)$ provides a refined value for $\beta$. Similarly, Equation \ref{eqn_2} allows us to obtain a new value for $\gamma$ from the straight-line fitting of $\log[\chi_0^{-1}(T)]$ versus $\log(\epsilon)$.
 During the fitting process, $T_c$ remains a variable parameter, adjusted to achieve the best fit. These new values of $\beta$, $\gamma$, and $T_c$ are used in Equation \ref{eqn_6} and the iterative process is continued until convergence is achieved. Following this procedure, optimal values $\beta=0.48$ and $\gamma=1.10$ are obtained for the present system. The resulting plots using these exponents show reasonably good parallel straight lines, as illustrated in Figure 7, and therefore, these exponents values are consistent with the critical behavior of the present system.
    \begin{figure}[!h]
	\centering
	\includegraphics[width=0.35\textheight]{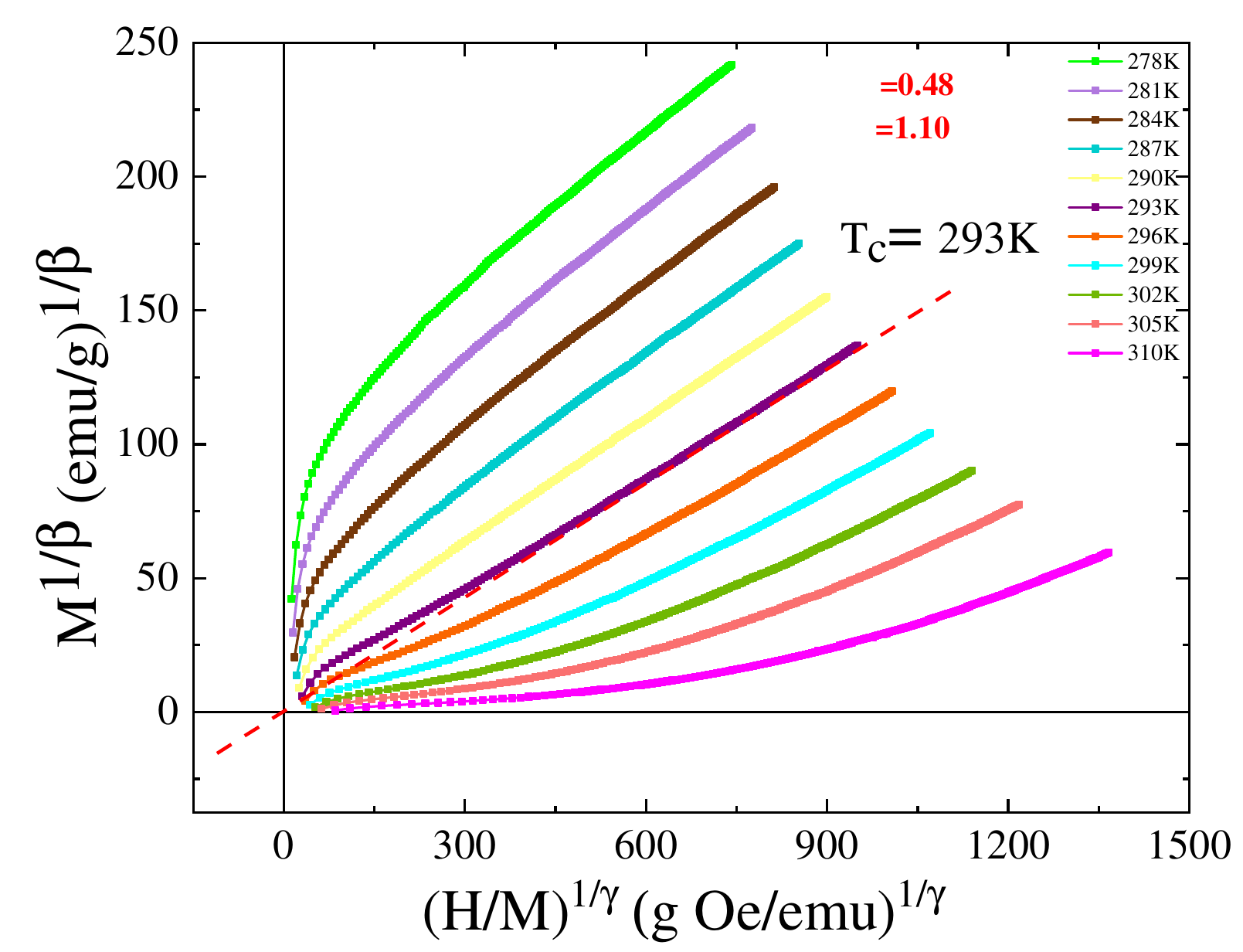}
	\caption{\label{Fig 4 MA Plot } Modified Arrott plot: $M^{1/\beta}$ vs. $(H/M)^{1/\gamma}$ isotherms with $\beta=0.48$ and $\gamma=1.10$. The dashed line highlights the isotherm at $T_C = 293$ K and passes through the origin.}
	\end{figure}
 At low field strengths, the lines exhibit curvature. This behavior is typical in many other complex oxide systems, as the magnetization represents a macroscopic average of many misaligned domains \cite{aharoni2000introduction, pramanik2009critical, triki2013unconventional, jia2019magnetism, chakraborty2023unusual}. At high field strengths, the isotherms align as nearly parallel straight lines. The extrapolated straight line from the linear region of the isotherm at 293 K passes through the origin, indicating that the corresponding temperature is consistent with previously determined $T_c = 293$ K.
 To verify the reliability of the critical exponents and $T_c$ associated with the modified Arrott plot, $M_S(T)$ and $\chi_0^{-1}(T)$ are extracted from Figure \ref{Fig 4 MA Plot } following the process mentioned above and plotted against temperature in Figure \ref{Fig 5 M vs Chi}. By fitting the $M_S(T)$ using Equation \ref{eqn_1} and $\chi_0^{-1}(T)$ with Equation \ref{eqn_2} yield $\beta=0.53\pm0.02$  with $T_c= 293.0\pm0.5$ K and $\gamma= 1.13\pm0.04$ with $T_c= 293.0\pm0.80$ respectively. These critical exponents and $T_c$ values closely match those derived from the modified Arrott plot in Figure \ref{Fig 4 MA Plot }. This consistency signifies successful convergence in the iterative approach, validating the reliability and accuracy of the obtained critical exponents.
    \begin{figure}[!h]
	\centering	\includegraphics[width=0.35\textheight]{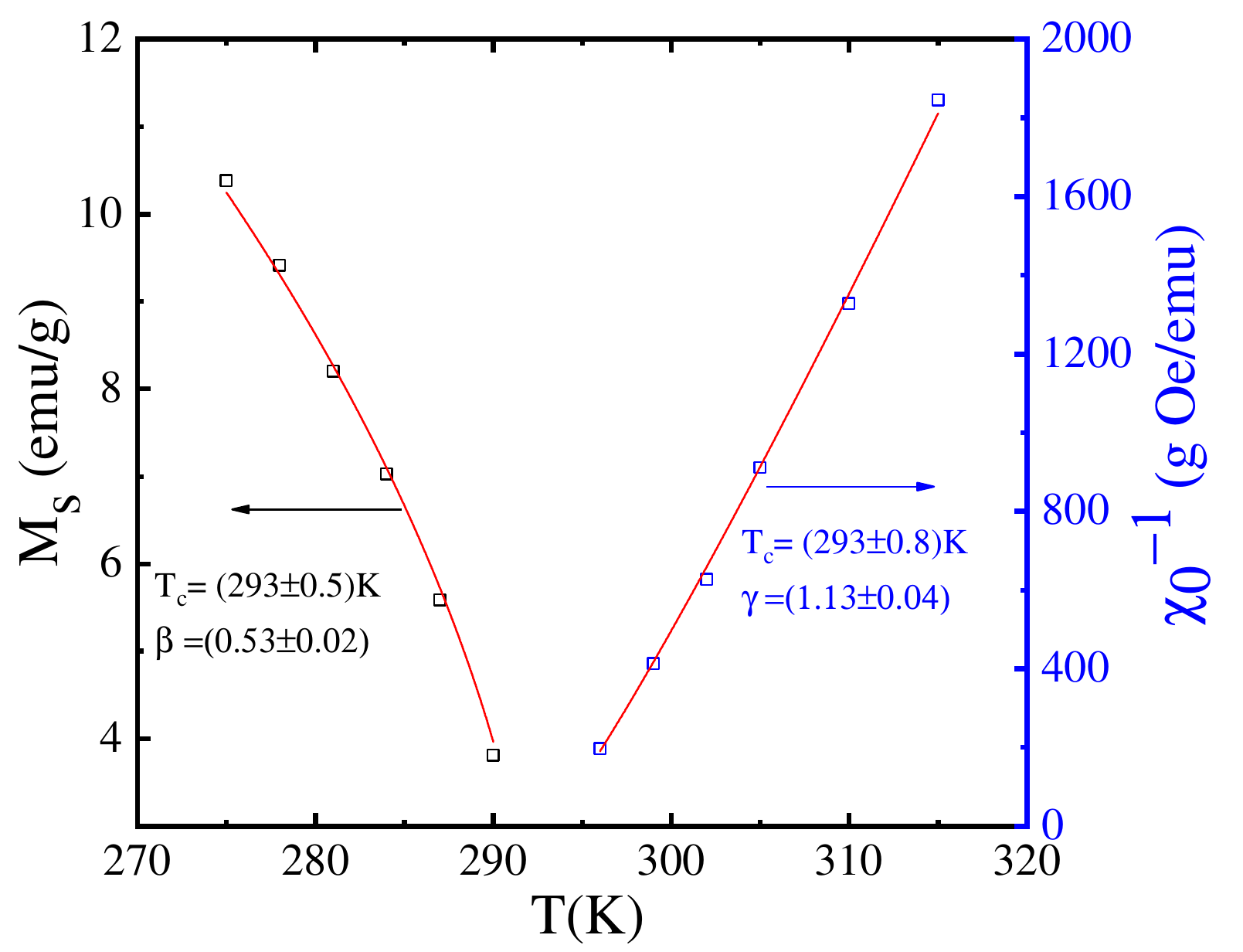}
	\caption{\label{Fig 5 M vs Chi} The temperature variation of the spontaneous magnetization $M_S(T)$ and inverse susceptibility $\chi_0^{-1}$ obtained from the extrapolation of modified Arrott plot in Figure \ref{Fig 4 MA Plot }. The values of $\beta$, $\gamma$ and $T_c$ are obtained from the straight line fittings of Equation \ref{eqn_1} and \ref{eqn_2}.}
	\end{figure}
 
\subsection{Kouvel-Fischer Plot}
\begin{figure}[!h]
    \centering
	\includegraphics[width=0.35\textheight]{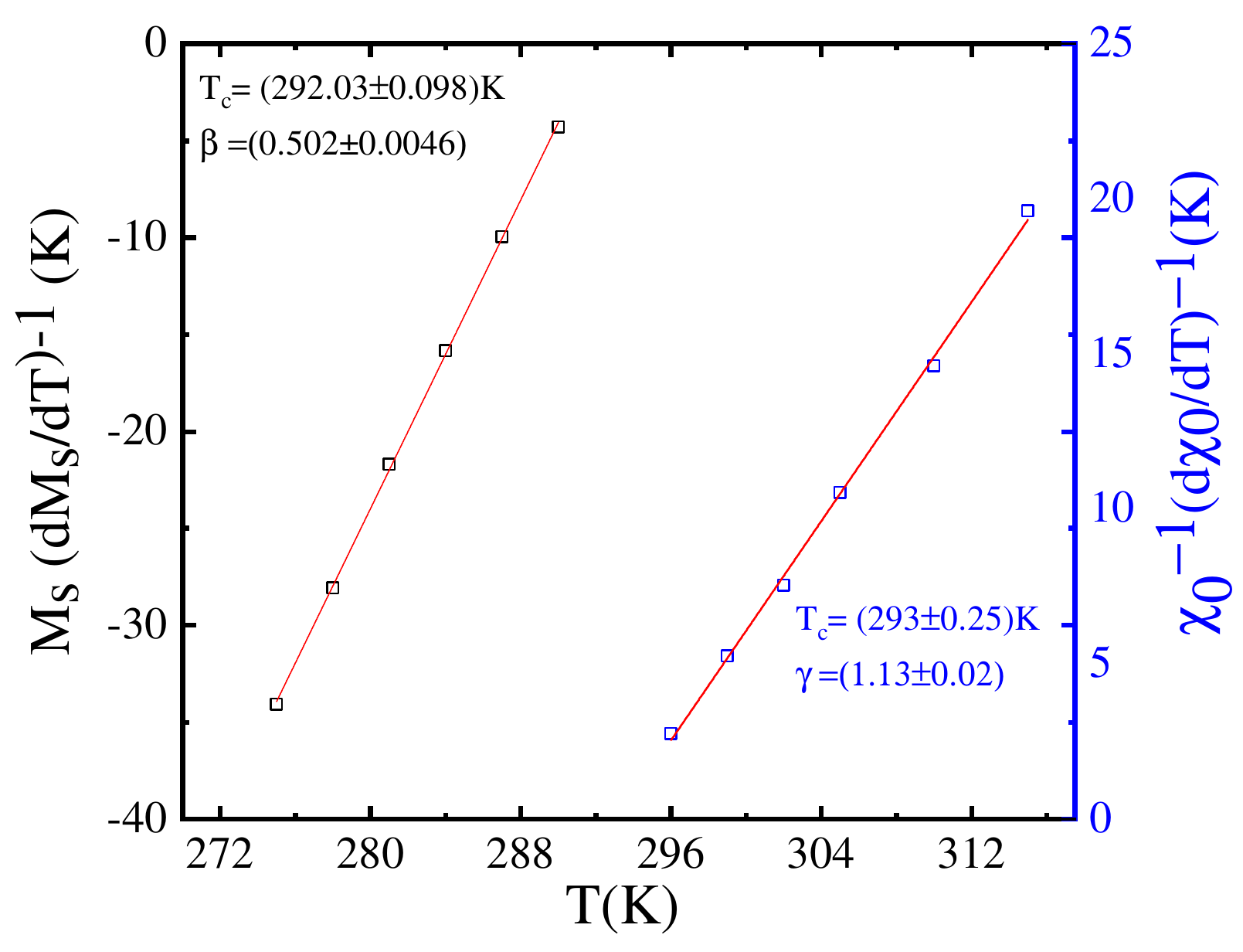}
	\caption{\label{Fig 6 Kouvel Fisher} Kouvel-Fischer plot of spontaneous magnetization $M_S$ and inverse susceptibility $\chi_0^{-1}$. Solid straight lines highlight the linear fit of the data.}
    \end{figure}

 In order to determine the critical exponents and  $T_c$ more accurately, the data of $M_S(T)$ and $\chi_0^{-1}(T)$ should be analyzed using the Kouvel-Fischer method \cite{kouvel1964detailed}:
	\begin{equation}
    M_S(T)\left[\frac{dM_S(T)}{dT}\right]^{-1} = \frac{(T-T_c)}{\beta}
	\label{kouvel-fischer-1}
	\end{equation}
	\begin{equation}
    \chi_0^{-1}(T)\left[\frac{d\chi_0^{-1}(T)}{dT}\right]^{-1} = \frac{(T-T_c)}{\gamma}
	\label{kouvel-fischer-2}
	\end{equation}
 According to these equations, plotting $M_S(T)[dM_S(T)/dT]^{-1}$ and $\chi_0^{-1}(T)[d\chi_0^{-1}(T)/dT]^{-1}$ against temperature $T$ should result in linear relationships. The slopes of these lines are inversely proportional to the critical exponents $\beta$ and $\gamma$, respectively. A significant advantage of the Kouvel-Fischer plot is that it enables the determination of $T_c$ without prior knowledge of its value. This is achieved by extrapolating the linear fits to their points of intersection with the temperature axis, as demonstrated in the Kouvel-Fischer plot for the (Ni$_{0.2}$Mg$_{0.2}$Co$_{0.2}$Cu$_{0.2}$Zn$_{0.2}$)(Mn$_{0.66}$Fe$_{0.66}$Cr$_{0.66}$)O$_{4}$ system in Figure \ref{Fig 6 Kouvel Fisher}.
 The analysis yields estimates for the critical exponents and transition temperature, namely, $\beta=0.502\pm0.004$ and $T_c=292.03\pm0.09$ K, based on the first fitting, and $\gamma=1.13\pm0.04$ and $T_c=293.08\pm0.80$ K, based on the second fitting. These values are consistent with previously obtained ones.

 \subsection{Critical isotherm exponent}
 \begin{figure}[!h]
    \centering
\includegraphics[width=0.35\textheight]{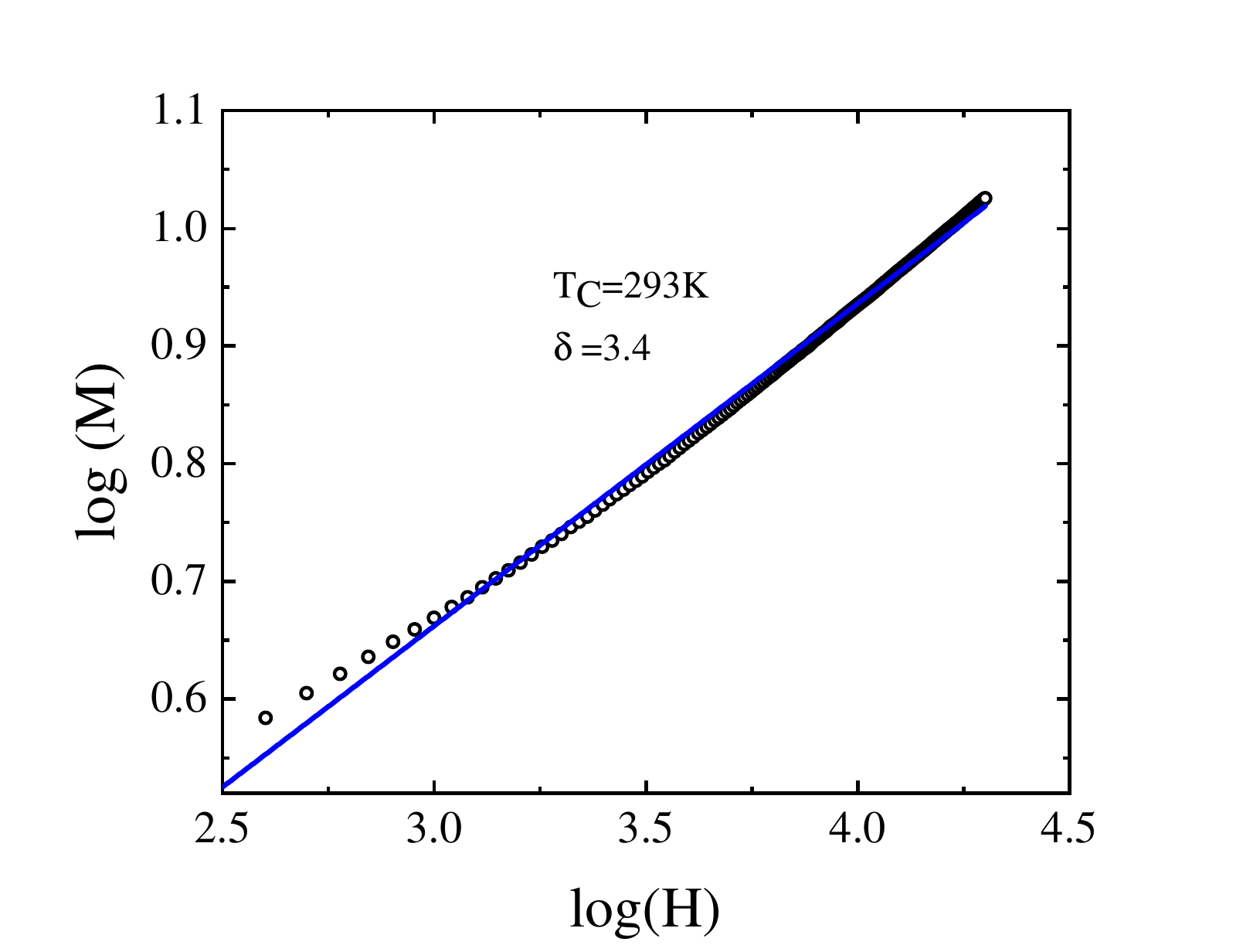}
	\caption{\label{Fig 7 log M vs Log H}
The log-log plot of isothermal magnetization (M) with applied magnetic field (H) for this high entropy oxide
collected at their respective T$_C$ = 293 K. The solid lines are the linear fits of critical isotherm relation described in Equation (3) to extract the $\delta$ value. }
\end{figure}
%%%%%%%%%%%%%%%%%%%%%%%%%%%%%%%%%%%%%%%%%%%%%%%%%%%%%%%%%%%%%%%%%%%%%%%%
The critical isotherm exponent, $\delta$, characterizes the power-law behavior of the isothermal magnetization near the critical point, providing comparisons among various systems with similar critical behavior. According to Equation \ref{eqn_3}, $\ln(M)$ versus $\ln(H)$ plot at the transition temperature should be a straight line with $\delta$ being its slope. For the present system, the high-field region yields $\delta=3.43\pm0.02$ as shown in Figure \ref{Fig 7 log M vs Log H}. Additionally, the Widom theorem relates \cite{widom1964degree} $\delta$ to $\beta$ and $\gamma$ through Equation  
    \begin{equation}
	\delta=1+\frac{\gamma}{\beta}
	\label{widom}
	\end{equation}
 Using the values of $\beta$ and $\gamma$ from the modified Arrott and Kouvel-Fischer plots, we obtain $\delta=3.29\pm0.006$ and $3.25\pm0.009$, respectively. The consistency of these values, obtained through different methods, confirms that the system obeys Widom scaling and demonstrates the accuracy of the estimated exponents ($\beta$ and $\gamma$) within experimental precision.

 \subsection{Conformity of scaling law}
 The critical exponents obtained through various methods are summarized in Table {\ref{Table2}}  alongside their theoretical values predicted for different universality classes.
\begin{table*} 
\caption{Values of critical exponents $\beta$ , $\gamma$ and $\delta$ as obtained from the modified Arrott plot, Kouvel–Fisher plot, and the critical isotherm are
given for {(Ni$_{0.2}$Mg$_{0.2}$Co$_{0.2}$Cu$_{0.2}$Zn$_{0.2}$)(Mn$_{0.66}$Fe$_{0.66}$Cr$_{0.66}$)O$_{4}$}.  Critical exponents corresponding to various universality classes are also given for comparison. }
\centering
\renewcommand{\arraystretch}{1.5}
\tabcolsep 2.5mm 
\begin{tabular}{|c|c|c|c|c|c|}     \hline\hline
System or Models & Technique & $\beta$ & $\gamma$ & $\delta$  &       References  \\ \hline
 HEO & Modified Arrott plot & 0.48 $\pm$ 0.01 & 1.10 $\pm$0.01 & 3.29$\pm$0.02 &  This work\\ \hline
 HEO & Kouvel–Fisher & 0.50$\pm$ 0.02 & 1.13 $\pm$0.01 & 3.26$\pm$0.03 &  This work\\ \hline
                                             
Mean field model Theory & Theory & 0.5& 1.0& 3.0&  \cite{kaul1985static}\\ \hline
3D-Ising model Theory & Theory & 0.325 &1.241& 4.82& \cite{kaul1985static}   \\ \hline
3D-Heisenberg model Theory & Theory & 0.365 & 1.386 &4.80&\cite {kaul1985static}\\\hline
Tricritical mean field model & Theory &0.25 &1 & 5& \cite{kaul1985static} \\ \hline

\end{tabular}
\label{Table1}
\end{table*}
%%%%%%%%%%%%

The comparison shows that the critical exponents obtained through various methods match closely with the mean field model amongst various conventional universality classes. It is crucial to further validate the consistency of the critical analysis and the obtained critical exponents by examining the universal scaling behavior, as described by Equation \ref{eqn_5}. 
    \begin{figure}[!h]
	\centering
\includegraphics[width=0.35\textheight]{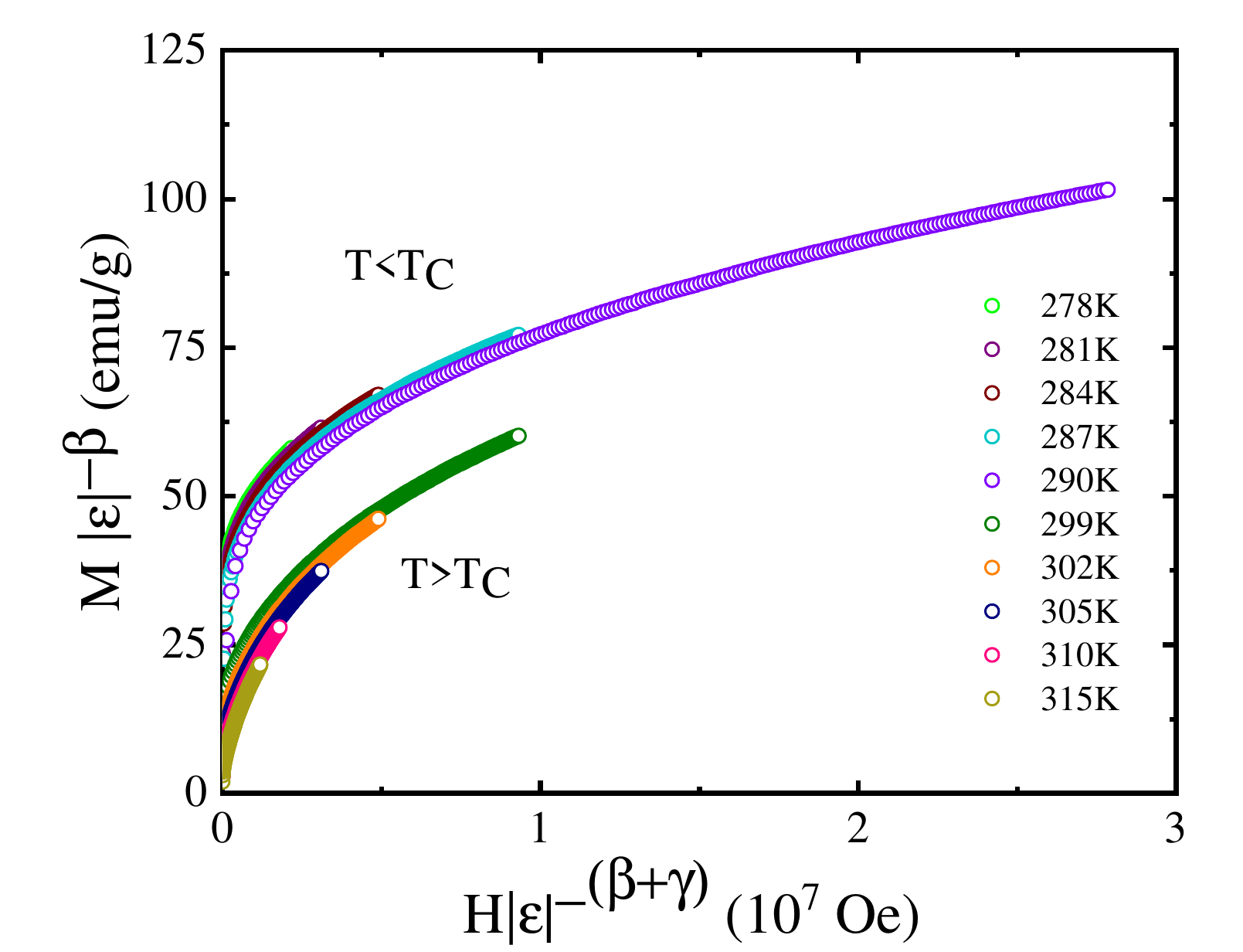}
	\caption{\label{Fig 9 RN} Scaling plots (renormalized magnetization $m\equiv\epsilon^{-\beta}M(H,\epsilon)$ vs. renormalized field $h\equiv\epsilon^{-(\beta+\gamma)}H$) below and above $T_c$. Values of $\beta$ and $\gamma$ that were determined from the modified Arrott plot are used.}
	\end{figure}
 The plot of the renormalized magnetization ($m$) against the renormalized field ($h$), as obtained using the values of $\beta$ and $\gamma$ obtained from the modified Arrott plots, is shown in Figure \ref{Fig 9 RN}.
 Notably, all data points converge into two distinct curves, one below $T_c$ and the other above $T_c$. This data collapse evidences the presence of a renormalized interaction within the critical regime. Renormalization interaction refers to the collective impact of fluctuations and interactions among the order parameters within the system, which becomes increasingly significant as the critical point is approached. This enhanced interaction modifies the behavior of the system, leading to the observed data collapse and demonstrating the existence of universal scaling laws near the critical point. Moreover, the confirmation of data collapse validates the estimated critical exponents and $T_C$.

 \subsection{Interpretation of the critical exponents from high entropy aspect}
 It is concluded from various analysis of this study that $\beta =0.48$, $\gamma=1.1$, $\delta=3.3$ and $T_C = 293$ K is a consistent set of critical exponents. This closely matches with the exponents of the conventional mean field model. This is a noteworthy result of the system. It is worth mentioning here that  critical behavior of many complex oxide materials (non-high entropy) with novel magnetism and functional properties, such as various perovskite, double perovskite, spinel and their derivatives have been extensively investigated \cite{PhysRevLett.101.077206,borah2020magnetocaloric,xu2016room,singh2020critical,yadav2021unconventional}. Very often, it is a common observation that their critical exponents are unique and do not belong to any of the conventional universality classes. There are many factors which causes the phase transition unconventional in these systems. Few well accepted facts are- inevitable oxygen vacancies, cluster formation and chemical homogeneity etc lading to mixed valence of the transition metals and resulting in distribution of exchange interaction strengths {\cite{chakraborty2023unusual}. From this perspective, it is remarkable to note that despite being a high entropy oxide—a complex system with multiple transition metal elements in both A and B sites—our present system adheres to the mean field model. This model simplifies the understanding of complex many-body systems by averaging the effects of all other particles on any given particle.
 In magnetism, it does not treat all spins individually. It Substitutes the individual spin - spin interaction between electrons by the interaction of the spins with an average or effective interaction which is basically the internal magnetic field proposed in Weiss molecular field theory of ferromagntism. Though this is an oversimplified assumption, it is interesting to observe that this picture holds consistently in such a complex system with numerous magnetic atoms with diverse spin moments in A and B sites. This is however, not very uncommon in case of cubic spinel systems.
Experimental observations in many spinel oxides yield critical exponents close to the mean field predictions, suggesting the validity of mean field theory in describing their magnetic behavior. A list of such spinel systems consistent with mean field model is given in Table III.
\begin{table*}
\caption{Spinel systems where critical exponents are consistent with the mean-field model}
\centering
\renewcommand{\arraystretch}{1.15}
\tabcolsep 4.5mm 
\begin{tabular}{|c|c|c|c|c|c|c|}     \hline\hline
Sr No & Spinel Oxide & $\beta$ & $\gamma$ & $\delta$  &  Theory   &  References  \\ \hline
1 & CoGa$_{1.2}$Fe$_{0.8}$ O$_{4}$ & 0.49  & 1.00 & 3.01 & Close to MFT & \cite {hu2022investigations} \\ 
&Ferromagnetic (T$_{C}$=236K)&   &  &  &  & \\ \hline
2 & Co$_{1+x}$Cr$_{x}$Fe$_{2-x}$ O$_{4}$ & 0.44 & 0.99 & 3.21 & Close to MFT & \cite{islam2023magnetic}\\ 
&Ferrimagnetic (T$_{C}$=636K)&   &  &  &  & \\ \hline
3 & Ni$_{0.6}$Cd$_{0.2}$Cu$_{0.2}$ Fe$_{2}$O$_{4}$ & 0.44 & 1.03 & 3.31 & Close to MFT & \cite{kouki2019microstructural} \\ 
&Ferromagnetic (T$_{C}$=655K)&   &  &  &  & \\ \hline

4 & Mn$_{0.5}$Zn$_{0.5}$Fe$_{2}$O$_{4}$ & 0.48 & 1.07 & 3.34 & Close to MFT & \cite {anwar2022investigation}\\ 
&Ferromagnetic (T$_{C}$=345K)&   &  &  &  & \\ \hline
\end{tabular}
\label{Table2}
\end{table*}
%%%%%%%%%%%%
Possible reasons for these systems to follow mean field model might be their high symmetric cubic crystal structure, the nature of superexchange interactions, and the uniform distribution of cations which make it possible that the assumptions of mean field theory are well-satisfied in these materials. Our high entropy spinel oxide system (Ni$_{0.2}$Mg$_{0.2}$Co$_{0.2}$Cu$_{0.2}$Zn$_{0.2}$)(Mn$_{0.66}$Fe$_{0.66}$Cr$_{0.66}$)O$_{4}$  is also based on spinel with cubic structure. While a similar rationale
might reasonably apply to our system, the distinct high-entropy nature of our system suggests
that entropy could also play a crucial role in our case. It is possible that the high entropy leads to a homogeneous distribution of multiple cations, as it is energetically favourable. This ensures that averaging of the local magnetic environments (also observed in 10 K NPD refinement) is a reasonably good approximation of the system and make it favourable to satisfy the mean field theory.

\section{Conclusions}
In conclusion, We have stabilized and studied a new high entropy spinel oxide with composition (Ni$_{0.2}$Mg$_{0.2}$Co$_{0.2}$Cu$_{0.2}$Zn$_{0.2}$)(Mn$_{0.66}$Fe$_{0.66}$Cr$_{0.66}$)O$_{4}$. The comprehensive structural (using NPD and X-ray diffrration), microstructural, magnetic, and X-ray absorption spectroscopy analyses reveal that despite the high degree of disorder, this oxide exhibits long-range ferrimagnetic ordering starting at 293 K. The material's impressive high saturation magnetization of 766 emu cm$^{-3}$, low coercivity of 100 Oe, high transition temperature around room temperature, and outstanding resistivity of 4000 ohm-cm at room temperature underscore its potential for high-density memory devices. The NPD analysis and RT-XAS measurements highlight the actual chemical composition of the materials as (Ni$_{0.1}$Mn$_{0.1}$Mg$_{0.2}$Co$_{0.2}$Cu$_{0.2}$Zn$_{0.2}$)(Mn$_{0.56}$Ni$_{0.1}$Fe$_{0.66}$Cr$_{0.66}$)O$_{4}$. Our investigation on the magnetic structure, using a collinear-type ferrimagnetic model with a propagation vector k = 0,0,0, along with various analytical techniques such as modified Arrott plots, Kouvel-Fischer analysis, and critical isotherm analysis, confirms a second-order phase transition. The critical exponents obtained are consistent with the mean field model, demonstrating that high entropy in this material leads to a homogeneous distribution of multiple cations, thereby validating the approximation of averaging local magnetic environments and supporting the mean field theory. We would like to emphasize that this is the only work on the critical behavior around the phase transition temperature on a complex high entropy oxide. This study not only highlights the potential of incorporating a diverse range of elemental components on both the crystallographic sites of a complex spinel oxide framework but also highlights the excellent opportunities for tailoring and enhancing the properties of strongly correlated materials.

\section{Acknowledgments}
S. M. acknowledges the SERB, Government of India, for the (SRG/2021/001993) Start-up Research Grant and UGC DAE CSR for the CRS project (project No. - CRS/2023-24/01/1001). We acknowledge the UGC-DAE Consortium for Scientific Research (CSR), Indore for the magnetization and XAS measurements. We would like to thank Dr. Rajeev Rawat, Dr. Mukul Gupta, Mr. Rakesh Sah, Mr. Avinash Wadikar, Mr. Kranti Sharma and Ms. Soumya Shephalika for their support during the magnetic and XAS measurements. UGC-DAE CSR, Mumbai is also acknowledged for the Neutron diffraction experiments.

%\section{References}
\bibliographystyle{apsrev4-2}
\bibliography{main.bib}

\end{document}